\documentstyle[12pt]{article}
\begin{document} 

\begin{titlepage}
\begin{flushright}
UA-NPPS 15/96 \\
hep-th/xxxxxxx \\
July 1996\\
\end{flushright}
\vspace*{0.2cm}
\begin{center}
\bf{BACKREACTION EFFECT IN THE} \\
\bf{TWO-DIMENSIONAL DILATON GRAVITY}\\
\vspace*{0.5cm}
C. Chiou-Lahanas,  G.A. Diamandis$^{1}$, B.C. Georgalas,\\
 A. Kapella-Economou and X.N. Maintas.\\
\vspace*{0.3cm}

{\it University of Athens. Physics Department}\\
{\it Nuclear and Particle Physics section}\\
{\it Panepistioupolis 157-71. Athens - Greece.}\\

\vspace*{0.5cm}
\bf{Abstract}\\
\end{center}  
\noindent
    In this work we find static black hole solutions in the context
    of the two-dimensional dilaton gravity, which is modified by the addition
    of an $R^2$ term. This term arises from the one-loop effective action
    of a massive scalar field in its large mass approximation.
    The basic feature of this term is that it does not
    contribute to the Hawking radiation of the classical black hole
    backgrounds of the model. From this point of view a class of the 
    solutions derived are interpreted as describing backreaction effects. In 
    particular it is argued that evolution of a black hole via non-thermal 
    signals is possible. Nevertheless this evolution seems to be 'soft', in 
    the sence that it does not lead to the evaporation of a black hole, 
    leaving the Hawking radiation as the dominant mechanism for this process. 
\vspace*{0.5cm}
\noindent
\vfill
\hrule
\vspace*{0.2cm}
\begin{flushleft}
{\it e-mail}: $^{1}$  gdiamant@atlas.uoa.gr
\end{flushleft}
\end{titlepage}
\newpage
 
\section{Introduction}
After the initial observation that pure two-dimensional gravity accepts 
black hole  solutions \cite{witten}, a
model was proposed describing the formation of a black hole by the  
collapse of conformal matter followed by the subsequent emission of 
Hawking radiation \cite{cghs}. 
Furthermore  the quantum evolution of the classical black hole background 
has been studied \cite{rst} - \cite{gidstr}. 
Thus two-dimensional dilaton  gravity has been recognized as a usefull
theoretical laboratory for the study of the quantum aspects of gravity,
and considerable activity persists in the recent years in this field.

In a recent publication \cite{chiou} the contribution of a 
massive scalar field
to the Hawking radiation rate of a static black hole in the context of 
the two-dimensional dilaton gravity, was calculated. The calculation shows
that the Hawking effect is enhanced in this case, in comparison with that of
the conformal
matter \cite{cghs}, provided that the mass of the quantized field is less 
than one, in units of the cosmological
constant of the model. Note that this happens to be the case if the 
scalar field
is the tachyon field. The tachyon field is an ordinary massive field in
two dimensions as is dictated from the stability requirement of the classical
background solutions. It was also recognized in \cite{chiou} that the thermal 
radiation effect is due to the coupling of the scalar field to the dilaton. 
In particular it 
was shown that terms of the form $R\beta(\Box)R$, in the effective action 
do not contribute 
to the Hawking radiation effect. This observation makes interesting the 
investigation of the evolution of a black hole in the presence of such purely 
geometric terms in the effective action. The effect of such terms is 
non thermal and deviation from the thermal evolution yields the possibility 
of retrieving information from the  
black hole. Indications about such deviations are already pointed out
in the context of the RST model \cite{rst,keski}.

In this work we address this problem, that is we try to get information
about the evolution of a black hole in the presence of such terms. In 
section 2 we set up the problem justifying that a simple term of the
form $R^2$ can emerge from the one-loop action in a large mass approximation.
In section 3 we derive the static black hole solutions of the two-dimensional
dilaton gravity after the addition of this term. It is found that the black 
hole geometries
which tend asymptotically to flat space with linear dilaton, can be 
associated in a certain way to the classical black hole backgrounds. It is
argued therefore that these solutions incorporate the backreaction effects
due to the term added in the Lagrangian.
Finally in section 4 we discuss our results, supporting further the 
above argument.
\section{Preliminaries}

The classical action of a scalar field coupled to the two-dimensional 
dilaton gravity is
\begin{equation}
S_{cl}=\frac{1}{2 \pi} \int d^2x \sqrt{-g} e^{-2 \phi} \left\{  \left[
R + 4 (\bigtriangledown \phi)^2 + 4 \lambda ^2 \right] -  
\left[(\bigtriangledown T)^2 -  m_0^2 T^2\right]
\right\}
\label{sclas}
\end{equation}
where $m_0^2$ is negative if the scalar field is the tachyon field, and
positive in the case of an ordinary massive field. Note that if we rescale
the scalar field as $T=e^{-\phi} \widetilde T$, then the field
$\widetilde T$ which has canonical kinetic terms is coupled to the 
dilaton through the quadratic coupling $-(Q + m^2) \widetilde T ^{2}$
where $m^2=\lambda^2+m_0^2$, $
Q=(\bigtriangledown\phi)^2 - \Box\phi - \lambda^2$.
In the case of flat space with linear dilaton, $Q=0$, while stability 
of the background, under scalar field perturbations,
enforces $m^2$ to be positive. The positivity of $m^2$ is also required
from the stability of the black hole backgrounds. Thus in two dimensions
the tachyon field is an ordinary propagating mode.

The non-local part of the one-loop effective action (which is responsible 
for the Hawking radiation), if we keep only quadratic terms in the background
fields $Q$, and $R$, is given by \cite{chiou,avramidi}
\begin{equation}
\Gamma_{nloc} = \frac{1}{8 \pi} \int d^2x \sqrt{g} \left\{ 
Q \beta^{(1)}(\Box)Q - 2Q \beta^{(2)}(\Box)R + R \beta^{(3)}(\Box)R \right\}  
\label{nloc}
\end{equation}
where
\begin{eqnarray}
\beta^{(1)}(\Box) &=& \frac{1}{\Box} \sqrt \gamma ln \frac{1+ \sqrt{\gamma}}{
1- \sqrt{\gamma}} \label{betao} \\
\beta^{(2)}(\Box) &=& \frac{1}{\Box} \left\{ \frac{\gamma - 1}{4 
\sqrt{\gamma}} ln \frac{1+ \sqrt \gamma}{1- \sqrt{\gamma}}  +
\frac{1}{2} \right\} \label{betat} \\
\beta^{(3)}(\Box) &=& \frac{1}{48 \Box} \left\{
\left(3-\frac{6}{\gamma} + \frac{3}{\gamma^2} \right) \sqrt{\gamma}
ln \frac{1+ \sqrt \gamma}{1- \sqrt{\gamma}} - \frac{6}{\gamma} +
10 \right\} \label{betath}
\end{eqnarray}
and $\gamma = \frac{1}{1 - \frac{4 m^2}{\Box}}$.
\newline
It was pointed out in \cite{chiou} that the purely geometric term does not 
contribute to 
the thermal radiation of the classical static black hole background, 
provided that the mass of the field is nonzero.  
Large mass expansion of this term yields
\begin{equation}
R \beta^{(3)}(\Box)R= -\frac{1}{60 m^2}R^2 + O(\frac{R^4}{m^4})
\label{expan}
\end{equation}
Thus if we consider the action
\begin{equation}
S = S_{cl} - k\int\sqrt{-g}R^2,
\label{action}
\end{equation}
with $k = 1/(240 m^2)$,
the second term can be thought of as a quantum correction in 
the above described 
approximation. The full nonlocal one-loop action, even in this  
approximation, contains also
terms like $Q^2$ and $QR$. These terms are responsible for the Hawking 
radiation of the black hole, as was shown in \cite{chiou}. 
In the present work, since we are interested
in isolating possible effects of non-thermal character, we ignore these 
terms.
\newline
The equations of motion for the action in (\ref{action}) are
\begin{eqnarray}
\Box \phi - (\bigtriangledown \phi)^2 + \frac{1}{4}R + \lambda^2 -
\frac{1}{4} (\bigtriangledown T)^2 - \frac{m_0^2}{4} T^2 &=& 0 
\label{dil} \\
\Box T -2 (\bigtriangledown \phi) (\bigtriangledown T) - m_0^2 T &=& 0  
\label{tach} \\
 e^{-2 \phi} \left\{ - \frac{1}{4} g^{\mu \nu} R 
- \phi ^{; \mu \nu} + \frac{1}{2} T^{, \mu} T^{, \nu} \right\} 
&-& \nonumber \\ 
k \left\{- g^{\mu \nu} ( \frac{1}{4} R^2 + \Box R) + R^{; \mu \nu} 
\right\} &=& 0 \label{metr}
\end{eqnarray}
We will seek static black hole solutions, with trivial scalar field 
configuration $(T = 0)$.
The resulting equations are insensitive in the mass of the scalar field, since
a change in $k$ can be compensated by a constant shift of the dilaton. Thus in 
the 
analysis that follows we take $k=1$. 
Working in the Schwarzchild gauge
\begin{equation}
ds^2 = -gdt^2 + \frac{1}{g}dr^2
\label{schw}
\end{equation}
the field equations take the form
\begin{eqnarray}
 g \phi^{''} - \frac{1}{4} g^{''} + g^{'} 
 \phi^{'} - g(\phi^{'})^{2} + 1  & = &  0, \label{sdil} \\
g^{''''} - e^{-2 \phi} \phi ^{''}  & = &  0, \label{smetr} \\
2 g g^{''''} - \frac{1}{2} (g^{''})^{2} + g^{'} g^{'''} + e^{-2 \phi} (
-\frac{g^{''}}{2} + g^{'} \phi^{'})  & = &  0. \label{smetr1}
\end{eqnarray}
where the derivations are with respect to the dimensionless variable 
$x = \lambda r$. The equation (\ref{sdil}) is the dilaton equation and 
(\ref{smetr}, \ref{smetr1}) are derived  from
the equations for the metric components (\ref{metr}) with the use of
(\ref{dil}).
We are interested in solutions which are asymptotically flat. In the gauge 
we are working 
$R(r) = -\lambda^2g''$.
So the requirement for asymptotically vanishing curvature 
gives two qualitatively different asymptotic
forms for the metric g:
$$
\begin{array}{cccccccc}
  &g& \rightarrow & 1 &, &as& & x\ \rightarrow \infty, \\
or  &g& \rightarrow & Ax + B & (A>0), &as& & x\ \rightarrow \infty, 
\end{array}
$$
where the condition $A>0$ is required by causality.
In the next section we will examine the two cases and derive the black 
hole solutions in each one.
\section{Black Hole Solutions}
\subsection{Case with asymptotically constant metric}

 A full analytical solution, in closed form, of the nonlinear system of the 
 equations of 
motion is not possible. Attempting to get numerical solutions with the  
desired properties is not straightforward because one has to fix six
initial values for the fields (and their derivatives). Even in such an
effort
there are not  physical criteria for detetmining the relative values of
the six initial conditions. The method we adopt is partially analytic
and partially numerical. In particular the solution of system is written in 
an iteration form. Althought it is not possible to get the solutions in 
closed form the iteration procedure permits a full control them. 
To be more precise, taking the solution at the first iteration step and
consequently using the values of the fields at an appropriate point as
initial conditions, we get full numerical solutions of the system of 
equations which respect the requirement of asymptotic linear dilaton
vacuum. Fufthermore this method has the 
advantage that permits exact determination of the ADM mass of the
corresponding black holes. In the rest of the subsection we will scetch
the method adopted and present the basic features of the solutions that 
we get.

With the required asymptotic behaviour of the metric, we can see from
the dilaton equation that the corresponding asymptotic form of the 
dilaton field is just linear as expected, that is $\phi_{as} = -x$.
Taking into account the asymptotic behaviour of the fields $g$ and $\phi$ 
we make the following ansatz

\begin{eqnarray}
g  &=&  e^{y(x)}, \nonumber \\ 
\phi &=& -x + h(x).
\label{ansatz}
\end{eqnarray}

Therefore we seek solutions with
$$
\begin{array}{ccccccc}
y(x), & h(x)&  & \rightarrow & 0, \\ 
 as   & x & & \rightarrow & \infty.
\label{asympt}
\end{array}
$$
Note that the black hole solutions arising from the classical action
(\ref{sclas}), with trivial scalar field configuration, are given by

\begin{eqnarray}
 y^c(x) &=& ln(1-Me^{-2x}), \nonumber \\
 h^c(x) &=& 0. 
\label{ldv} 
\end{eqnarray}

Considering the equations (\ref{sdil}) and (\ref{smetr1}) 
the corresponding first order system takes the form
\begin{equation}
\frac{d \overrightarrow{{\bf X}}} {dx} = {\bf A}(x) \overrightarrow{{\bf X}} +
\overrightarrow{{\bf F}}(x,\overrightarrow{{\bf X}})
\label{first}
\end{equation}
where
\begin{eqnarray}
\overrightarrow{{\bf X}} &=& (h_0=h, h_1=h^{'}, 
\widehat{y}_3 = -e^{-2x}y^{'''},
y_2 = y^{''}, y_1 = y^{'}, y_0 = y), \nonumber \\
\overrightarrow{{\bf F}}(x, \overrightarrow{{\bf X}}) &=& 
(0, F, G, 0, 0, 0).
\label{explan}
\end{eqnarray}
The matrix
\begin{equation}
{\bf A} = \left[
\matrix{0 & 1 & 0 & 0 & 0 & 0\cr
        0 & -2 & 0 & 1/4 & 1 & 1\cr
        0 & 0 & -2 & -1/4 & -1/2 & 0\cr
        0 & 0 & -e^{2x} & 0 & 0 & 0\cr
        0 & 0 & 0 & 1 & 0 & 0\cr
        0 & 0 & 0 & 0 & 1 & 0\cr}
        \right]
\end{equation}
determines the linear part of the system (\ref{first}). The nonlinearity is 
hidden in  
the functions $F$ and $G$, in (\ref{explan}), with
\begin{eqnarray*}
F &=& (1-y_0-e^{-y_0}) + h_1^{2} + h_1 y_1 + \frac{y_1^{2}}{4}, \\
G &=& e^{-2x} \left(\frac{5}{4} y_1^{4} +\frac{11}{4} y_2^{2} +
7 y_1^{2} y_2 \right) + \frac{1}{4} \left(1-e^{-2h_0 - y_0}\right) \left(
y_2 + 2y_1 \right)  \nonumber \\
&+& \frac{1}{2}e^{-2h_0 - y_0}y_1h_1 - \frac{9}{2}
y_1 \widehat{y} _3.
\end{eqnarray*}
The eigenvalues of the matrix ${\bf A}$ are $(0, 0, -2, -2, 
-e^x/2, e^x/2)$. The three negative eigenvalues of this matrix yield
the solutions of the linear system which vanish asymptotically. Furthermore
one
expects that the solutions of the full system, having the desired 
asymptotic behaviour, depend on three parameters too.
The system (\ref{first}) can be written further in integral form 
\begin{equation}
\overrightarrow{{\bf X}}(x) = {\bf Y}(x) \overrightarrow{{\bf C}} +
{\bf Y}(x) \int_{x_0}^x
{\bf Y}^{(-1)}(t)
\overrightarrow{{\bf F}}(t,\overrightarrow{{\bf X}}(t))dt
\label{integr}
\end{equation}
where ${\bf Y}$ is a fundamental solution of the linear system
satisfying the equation
$$
\frac{d {\bf Y}(x)}{dx} = {\bf A}(x){\bf Y}(x)
$$
and $\overrightarrow{{\bf C}}$ are the corresponding constants of integration.
The above integral form dictates the iterative procedure. At each
step of the iteration the solution is
\begin{equation}
\overrightarrow{{\bf X}^{(n)}}(x) = \overrightarrow{{\bf X}^{(0)}}(x)  +
{\bf Y}(x) \int_{x_0}^x
{\bf Y}^{(-1)}(t)
\overrightarrow{{\bf F}}(t,\overrightarrow{{\bf X}^{(n-1)}}(t))dt
\label{itsol}
\end{equation}
where $\overrightarrow{{\bf X}^{(0)}}$ stands for the general 
solution of the
linear system which has the required asymptotic behaviour.
\newline
The linear part of the system (\ref{first}), as can be easily seen, 
is equivalent
to a system of two algebraic relations
\begin{eqnarray}
h_1(x) &=& be^{-2x} + \frac{1}{2}y_0(x) + \frac{1}{4}y_1(x), 
\label{algebr1} \\
\widehat{y} _3(x) &=& - \frac{a}{2} e^{-2x} - \frac{1}{4} y_1(x)
\label{algebr2}
\end{eqnarray}
and a second order differential equation
\begin{equation}
y_1^{''} - \frac{e^{2x}}{4} y_1 = \frac{a}{2}.
\label{difeq}
\end{equation}
Notice that the constants $a$, and $b$ are related to the double
eigenvalue (-2) of the matrix ${\bf A}$. The general solution of the
equation (\ref{difeq}) is
\begin{equation}
y_1(x)=d_1 K_0 (\frac{e^{x}}{2}) + d_2 I_0 (\frac{e^{x}}{2}) +
a\left[A(x) - \frac{x}{2} K_0 (\frac{e^{x}}{2}) \right],
\label{difsol}
\end{equation}
where $K_0$, $I_0$, are the modified Bessel functions of order zero and 
$A(x)$ is defined as
\begin{equation}
A(x) = \frac{1}{2} \left\{- I_0 (\frac{e^{x}}{2}) \int_{x}^{\infty} dt
K_0 (\frac{e^{t}}{2}) - K_0 (\frac{e^{x}}{2}) \int_{- \infty}^{x} dt 
 \left[I_0 (\frac{e^{x}}{2})  - 1 \right] \right\}.
\label{defal}
\end{equation}
From the asymptotic behaviour of the Bessel functions and their integrals
one can see that the parameters $d_1$ and $d_2$ correspond to the
eigenvalues $-e^{x}/2$ and $e^{x}/2$ of the matrix ${\bf A}$, respectively, 
and that
\begin{equation}
A(x) \sim -2 e^{-2x} + O(e^{-4x})
\label{albeh}
\end{equation}
All other fields can be derived from (\ref{difsol}) by differentiation or 
integration, 
or through the algebraic relations (\ref{algebr1}, \ref{algebr2}). 
Thus we conclude that the
solution  $\overrightarrow{{\bf X}^{(0)}}$ depends on three parameters
$a$, $b$ and $d_1$. Especially for the components $y_0$, $h_0$ and
$h_1$ we get the asymptotic behaviour
\begin{eqnarray}
y_0^{(0)}(x) &=& ae^{-2x} + O(e^{-4x}) \label{linmetr} \\
h_0^{(0)}(x) &=& \frac{-b}{2}e^{-2x} + O(e^{-4x})  \\
h_1^{(0)}(x) &=& be^{-2x} + O(e^{-4x}). \label{lindil} 
\end{eqnarray}
Taking also into account the asymptotic behaviour of the integral kernel
${\bf Y}(x) {\bf Y}^{-1}(t)$ of equation (\ref{itsol}), 
after some algebraic work,
one can verify that all the fields get contributions of order
$e^{-4x}$ at the first iteration step. Note that the above 
iterative method applied in
the case of the classical equations of motion yields the solution
in closed form

\begin{equation}
\label{alpha}
y_0(x) = \sum_{n=0}^{\infty} y_0^{(n)} = ln(1+ae^{-2x}),
\end{equation}
where $-a$ is the mass of the black hole.
\newline
In order to derive full solutions of the system we take the solution
at the first step of the iteration procedure. The corresponding expressions 
are not given here since they are quite complicated. Then we use the values 
of the
fields obtained in this way, at a point that they approach the corresponding
values of the flat space, as initial conditions and subsequently we solve
numerically the system. We indeed find solutions with the required
asymptotic behaviour. This is due to the stability of the iteration
ptocedure. The solution satisfy also the equation (\ref{smetr}) 
that was not used so far
verifying the compatibility of the full system of the equations of 
motion. Moreover these solutions for quite natural values of the parameters
(this notion will be explained later) exhibit black hole characteristics.
In particular they have an horizon at the point that the metric
function $g$ vanishes. In figures 1 and 2 we give the plots of the metric  
and the dilaton fields respectively for a typical structure coming out from 
the numerical solution of the system of the equations of motion.

Let us proceed by discussing the physical characteristics of the black holes
derived. All these black holes have the same temperature with the
classical solutions of the CGHS model \cite{cghs}
\begin{equation}
T = \frac{\lambda}{2 \pi}.
\label{temper}
\end{equation}
This holds since the corresponding backgrounds share the same asymptotic
behaviour, and can be easily seen from the periodicity of the 
compactified time
coordinate in Euclidean signature.  Another way to find the value of the
temperature is using the relation
\begin{equation}
T = \frac{\lambda}{4 \pi} \left \vert \frac{dg}{dx} \right \vert _ {
x=x_H},
\label{deftemp}
\end{equation}
where $x_H$ denotes the position of the horizon. The
assignment
$$
z = \int^{2x} \frac{1}{g(2x)}
$$
transforms the Schwarchzild gauge to the conformal one. In the  
conformal gauge the 
apparent horizon is determined from the value $1/2$ of the derivative
of the conformal factor. For the static geometries that we discuss 
here the apparent horizon coincides with the event horizon. Thus the
derivative in (\ref{deftemp}) calculated via its relation with the conformal 
gauge yields the value in (\ref{temper}).

The ADM mass of a static black hole, approaching asymptotically the
linear dilaton vacuum, is given by \cite{witten}
\begin{equation}
M = lim_{x \rightarrow \infty}\left\{2e^{2x}\left[h_1(x) + 
\frac{1}{2}(e^{-y_0(x)} - 1)\right] \right\}.
\label{masdef}
\end{equation}
This formula is exactly applicable in the case that we examine here.
For the classical solution, with the values of the fields $y_0$ and $h_1$ 
given in (\ref{ldv})
the above formula leads to the mass M (in units of the cosmological 
constant). In the quantum case we see that the linear part of the solution
is sufficient
to determine the black hole mass. Inserting in 
(\ref{masdef}) the forms of $h_1$ and $y_0$
given in  (\ref{linmetr}, \ref{lindil}) we find that
\begin{equation}
M = 2b - a.
\label{adm}
\end{equation}
The above result is exact since it is already pointed out (see also 
\ref{albeh}) that the
first iteration step and therefore all the steps in the iteration
procedure give corrections to the fields of order at least
$e^{-4x}$. These corrections do not contribute to the ADM mass. 
Thus we see that
all solutions derived in this way are of finite ADM energy. The natural
values of the parameters $a$ and $b$ refered above are those that keep
the ADM mass non-negative.

The solutions of the classical equations of motion, with trivial scalar field
background, given in (\ref{alpha}) depend only on the parameter $a$. 
Negative values of $a$ give black hole solutions with mass $-a$ while 
$a=0$ is just the flat space (in all these cases the dilaton is linear).
In the presence of the $R^2$ term we have derived solutions that depend on
two more parameters ($b$ and $d_1$). From the triparametric family of 
solutions the class with $b=d_1=0$ are corrections of the classical
geometry. These corrections do not alter the physical characteristics
of the black hole, as is seen from (\ref{temper},\ref{adm}). Furthemore
for given $a$ the presence of the quantum $R^2$ term, yields black hole
solutions with mass different from the mass of the corresponding classical
black hole ($b \not = 0$). We postpone the more detailed discussion about the 
interpretation of these solutions as backreaction effects for the next 
section.

Before closing this subsection let us note that the black hole solutions 
derived are stable against scalar field perturbations.  This can be seen 
by applying the method invoked in \cite{myung,tachyon}, for the 
study of the stability of the black hole solutions in the two-dimensional
dilaton gravity. In particular the equation satisfied by the space part of 
the scalar field, after separation of variables, is Schr\"odinger-like. 
Negative eigenvalues of the equation correspond to growing modes of
the scalar field perturbations indicating unstability of the background
geometry. In figure 3 the potential (U) of the eigenvalue equation is shown.
Clearly there are not  negative eigenvalues in this potential, justifying 
the assertion of stability. The form of the potential shows the 
repulsive character of the $R^2$ terms. 
Note that analogous solutions that have been found in \cite{kanti} four 
dimensions turned out to be unstable \cite{mavrom}.

\subsection{Case with asymptotically linear metric}
As in the previous subsection using the asymptotic form of the metric
we determine from the dilaton equation (\ref{sdil}) the asymptotic form of 
the dilaton
field, which reads
\begin{equation}
e^{-2\phi_{as}} = c_1I_0(\frac{1}{A}\sqrt{B+Ax}) +
c_2K_0(\frac{1}{A}\sqrt{B+Ax})
\label{asdil}
\end{equation}
The procedure adopted in the former case cannot be
followed easily here since an ansatz analogous to (\ref{ansatz})
is not obvious in this case. This makes the iteration procedure quite 
complicated.
Instead we proceed as follows. Using the asymptotic form of the dilaton
we integrate succesively the equation (\ref{smetr}). 
In this way we get more accurate forms of the
metric and its derivatives in the asymptotic region. The corresponding
values at some point are taken as the initial values for a numerical 
solution of the system.
We find that there exist solutions with the abovementioned asymptotic
behaviour, which exhibit the characteristics of a black hole. Namely
as in the previous case the function $g$ vanishes at some point
indicating thus the position of the horizon. 
Such kind of 
solutions, which  are 
asymptoticaly Rindlerian black holes, have already been found in the context
of other, different models of dilaton gravity in two dimensions 
\cite{mann,odintsov}. 
In general these black holes appear as the end point of gravitational 
collapse of localized matter. In our case the solutions derived cannot
be considered as backreaction effects since their asymptotic behaviour
is completely different from that of the classical black holes. In particular
besides the linear behaviour of the metric the solutions are strongly
coupled ($ \phi \rightarrow \infty $ as $x \rightarrow \infty$). 
As is seen from (\ref{smetr}, \ref{asdil}) only the choice $c_1 = 0$, lead 
to black hole solutions. In figure 4 we show the behaviour of the function
$g$ and the dilaton field in this case.  These  black hole solutions 
turn out to be stable against scalar field perturbations, as in the previous
case.  In general the stability of all the black hole solutions we have
found is among the characteristics of the two-dimensional consideration. 
As is pointed out in \cite{mavrom} this feature does not persists in
four dimensions.

\section{Discussion - Conclusions}

In this work we found the static black hole solutions, with asymptotically
vanishing curvature, in the context of two-dimenasional gravity, modified
by the presence of an $R^2$ term. This term is shown to arise from
the one-loop effective action of a massive scalar field, coupled to the 
theory, in its large mass expansion. The characteristic of this term is that
it does not contribute to the Hawking radiation of the classical
black hole background.

Two kinds of black hole solutions exist in this model differing in 
their asymptotic behaviour. The first describes symptotically Rindlerian
black holes. These solutions turn out to be strongly coupled, in the 
sense that the dilaton field diverges in the asymptotic region. There
is no connection with the classical black hole backgrounds of the 
two-dimensional dilaton gravity. So they are not considered to 
describe quantum correction or the end point of the evolution of a
classical black hole. 

The second class approaches asymptotically the linear dilaton vacuum. 
The solutions of this kind depend on three parameters ($a$, $b$ and
$d_1$). The parameter $a$ characterizes the classical solutions, determining
the ADM mass of the corresponding black hole geometries. From the new 
solutions, in the presence of the $R^2$ term, the subclass with $a<0$,
$b=d_1=0$ have the same physical characteristics of the associated
classical black hole. These solutions are backreacted ones. This backreaction
is compatible with the fact that the term added in the action does not
contribute to the Hawking radiation effect. That is for any classical
black hole background there is a new one of the quantum corrected action
with the same mass. The correction appears as a shift of the horizon
to larger values of $x$. In these solutions, as can be seen from the 
continuation of the metric behind the horizon of by transforming to the
conformal gauge, the physical singularity appears at finite value of $x$.
Moreover as is seen in figure 2  the dilaton ceases from being
monotonic in the vicinity of the horizon. This behaviour accounts for the
repulsive nature of the $R^2$ terms \cite{mavrom}, a feature that is also
shown in the plot of the potential (fig. 3) arising in the stability 
analysis.

Besides this class of solutions the presence of the $R^2$ term may turn
on the parameters $b$ and $d_1$ also. From these parameters, for given
$a$, the parameter $b$ represents the contribution of the dilaton field
to the ADM mass of the black hole (such a contribution is absent in the
classical case). The numerical analysis shows
that there are black hole solutions with either positive or negative $b$. 
Of course a complete scenario of the black hole formation and evolution, that
exists in the context of the RST model \cite{rst}, is missing in our
case. Nevertheless it is quite natural to assume that there exist
time dependent solutions interpolating between a classical or a backreacted
solution ($a<0$, $b=d_1=0$), and solutions with $b \not =0$ (more 
accurately $b=b(t)$, $d_1=d_1(t)$). Thus in this
model evolution of a black hole in the presence of a quantum term
which does not give the usual thermal effect can in principle be described.

From the evolution proccesses which may be described, under the above mild 
assumption, those with $b>0$ represent the absorption of energy by the 
black hole. The initial and final states have the same temperature 
(\ref{temper}) but
the mass of the final black hole is increased, according to (\ref{adm}).
Among the solutions with $b>0$ those with $a=0$ describe the 
formation of a black hole in the presence of the $R^2$ term.
The inverse proccess is more interesting, where the black hole may lose
part of its mass ($b<0$). This is achieved through non-thermal signals. 
Thus it is possible to get some information out of a black hole if the 
decreasing of its mass is due to the presence of terms of the kind considered
so far. The question that arises here is whether this evolution may be
responsible for a complete disappearance of a black hole, or differently 
how far can the mass of the black hole descent down in the presence of
the term $R^2$. The answer to this question seems to be that this evolution
is 'soft'.  The black hole may lose part of its energy but we cannot argue,
from our analysis, that  it can reach to a naked singularity. At first our
assumption of the time dependent solution interpolating between two static
ones is most likely to hold if the difference in the masses of the initial
and final configurations is not large ($ \vert b \vert \ll \vert a \vert $). 
On the other hand even
if we relax this condition we see that chosing $b = a/2$, we still
have a black hole structure with vanishing ADM mass. This means that the
model we consider accepts massless black holes. The interpretation
of structures with negative ADM energy (if we let $b$ to become more 
negative) is obscure to us. Of course in order to understand further 
the energy balance and therefore the permitted range of the parameter
$b$, (given the value of $a$) one has to consider solutions with
nontrivial scalar field configuration.

The entropy of the one-loop solutions that we have derived can be calculated 
by the method described in \cite{wald,myers}, althought the formula given
in \cite{myers} is a simple choice for the entropy, due to ambiguites
of the Wald construction in the presence of the $R^2$ term.
The result is that it is enhanced , on the event horizon, relative to the 
entorpy of the classical black hole solutions. 
This increase is mainly due to the behaviour of the dilaton field near the
horizon as is seen in figure 2. The increasing of the 
entropy is in agreement with the interpretation of the solutions as 
incoprorating backreaction effects. 

In general it is difficult to follow the evolution of a black hole in
this model. A better understanding could be achieved by studying the 
solutions with nontrivial scalar field, as is already refered. 
From our analytical consideration we see that the matrix {\bf A} in the
linear part of the system in (\ref{first}), extended to include
the equation of motion of the scalar field, possesses one more negative
eigenvalue, associated with T. Thus it is most likely that, as happens in the
classical case \cite{kostelecky,tachyon}, static solutions 
with non-trivial scalar field configuration exist. Furthermore the 
stability under scalar field perturbations indicate that time dependent 
solutions, which may appear in the evolution, exist also.
Besides this the consideration
of the terms $Q^2$ or $QR$, which appear in the one-loop effective action 
in the same approximation, will be helpfull.
These terms are responsible for the Hawking radiation of the classical
black hole background. Discrimination between the two cases could lead
in a better understanding of the quantum effects in the background
of a black hole.
\vspace*{1cm}
\newline

{\bf Acknowledgements}
\vspace*{0.3cm}
\newline

This work was partially supported by C.E.E. Science Program SCI-CT92-0792  
and by the 1995 PENED (No 512) program supported by the General Secretariat 
for the Research and Technology of the Hellenic Department of Research and
Technology.
We are gratefull to C. Bachas and N. Mavromatos for usefull discussions. The
numerical work is done with the use of MATHEMATICA.

\newpage
\noindent
{\bf Figure Captions}
\vspace*{1cm}
\newline

\parindent=-1cm
{\bf Fig. 1}  The behaviour of the metric for the black hole 
solutions that approach asymptotically the linear dilaton vacuum. 
The position of the horizon is shown at the point where the function
$g$ vanishes. The particular solution presented is determined by $a=-10$, 
$b=1$, $d_1=10$.
\vspace*{0.5cm} 
\newline

\parindent=-1cm   
{\bf Fig. 2} The behaviour of the dilaton field shows the inclination from 
linearity in the vicinity of the horizon.
\vspace*{0.5cm} 
\newline

\parindent=-1cm   
{\bf Fig. 3}  The potential arising in the stability analysis of the black 
hole solution presented in figure 1. The barrier form shown exludes negative 
eigenvalues of the corresponding Schr\"odinger equation. This means that a 
scalar field wave packet does not develop growing modes.
\vspace*{0.5cm} 
\newline

\parindent=-1cm 
{\bf Fig. 4}  The solid line shows the metric and the dashed
line the dilaton field for the asymptotically Rindlerian solutions.
The metric grows linearly while the dilaton behaves asymptotically as
$\sqrt{x}$.


\vspace*{1cm}
\begin{thebibliography}{99}

\bibitem{witten}
E.Witten, {\it Phys. Rev.} {\bf D44} (1991), 314.

\bibitem{cghs}
C.G. Callan, S.B. Giddings, J.A. Harvey and A. Strominger, 
{\it Phys. Rev.} {\bf D45} (1992), R1005.  

\bibitem{rst}
J.G. Russo, L. Susskind and L. Thorlacious, 
{\it Phys. Rev.} {\bf D46} (1992), 3444; 
{\it Phys. Rev.} {\bf D47} (1993), 533. 

\bibitem{lowe}
D.A. Lowe,  {\it Phys. Rev.} {\bf D47} (1993), 2446.

\bibitem{piran}
T. Piran and A. Strominger,  {\it Phys. Rev.} {\bf D48} (1993), 4729.  

\bibitem{bilcal}
A. Bilal and C. Callan, {\it Nucl. Phys.} {\bf B394} (1993), 73.

\bibitem{alwis}
S.P. de Alwis, {\it Phys. Lett.} {\bf B289} (1992), 330.

\bibitem{gidstr}
S.B. Giddings and A. Strominger, {\it Phys. Rev.} {\bf D46} (1993), 2454.

\bibitem{chiou}
C. Chiou-Lahanas, G.A. Diamandis, B.C. Georgalas, X.N. Maintas and E. 
Papantonopoulos, {\it Phys. Rev.} {\bf D52} (1995), 5877.

\bibitem{keski}
E. Keski-Vakkuri and S.D. Mathur, {\it Phys. Rev.} {\bf D50} (1994), 917.

\bibitem{avramidi}  
I.G. Avramidi, {\it Phys. Lett.} {\bf B236} (1990), 443.

\bibitem{myung}
Y.S. Myung,  {\it Phys. Lett.} {\bf B234} (1994), 29.

\bibitem{tachyon}
G.A. Diamandis, B.C. Georgalas and E. Papantonopoulos, 
{\it Mod. Phys. Lett.} {\bf A 10} (1995), 1277.

\bibitem{kanti}
P. Kanti and K. Tamvakis,  {\it Phys. Rev.} {\bf D52} (1995), 3506.

\bibitem{mavrom}
N.E. Mavromatos and E. Winstanley, {\it Phys. Rev.} {\bf D53} (1996), 3190. 

\bibitem{mann}
R.B. Mann,  {\it Phys. Rev.} {\bf D47} (1993), 4438.  

\bibitem{odintsov}
E. Elizalde, P. Fosalda-Vela, S. Naftulin and S.D. Odintsov, 
{\it Phys. Lett.} {\bf B352} (1995), 235.

\bibitem{wald}
R.M. Wald,  {\it Phys. Rev.} {\bf D48} (1993), 3427.

\bibitem{myers}
R.C. Myers,  {\it Phys. Rev.} {\bf D50} (1994), 6412. 

\bibitem{kostelecky}
V.A. Kostelecky and M.J. Perry, {\it Phys. Lett.} {\bf B322} (1994), 48.


\end{thebibliography}
\end{document}